\documentclass[prd,aps,groupedaddress,showpacs,preprintnumbers,
superscriptaddress,preprint,
nofootinbib]{revtex4}%
\usepackage{graphicx}
\usepackage{amsmath}
\usepackage{bm}
\usepackage{epsfig}
\usepackage{amsfonts}
\usepackage{multirow}
\usepackage{mathrsfs}
\usepackage{amssymb}
\usepackage{bbm}
\usepackage{slashbox}

\begin{document}
\vspace*{1cm}
\title{\mbox{}\\[10pt]
Relativistic Corrections to the Exclusive
Decays of \\
$\bm{C}$-even Bottomonia into $\bm{S}$-wave Charmonium Pairs}

\author{Wen-Long Sang}
\affiliation{Department of Physics, Korea University, Seoul 136-701, Korea}
\author{Reyima Rashidin}
\affiliation{Key Laboratory of Frontiers in
Theoretical Physics, The Institute of Theoretical Physics, Chinese
Academy of Sciences, Beijing 100190, People's Republic of China}
\affiliation{School of Physics Science and Technology, Xinjiang University,
Urumqi 830046, People's Republic of China}
\author{U-Rae Kim}
\affiliation{Department of Physics, Korea University, Seoul 136-701, Korea}
\author{Jungil Lee}
\affiliation{Department of Physics, Korea University, Seoul 136-701, Korea}
\affiliation{KISTI, Daejeon 305-806, Korea}

\date{\today}


\begin{abstract}
Within the nonrelativistic quantum chromodynamics (NRQCD) factorization
formalism, we compute the relativistic corrections to the exclusive
decays of bottomonia with even charge conjugation parity into
$S$-wave charmonium pairs at leading order in the strong coupling constant.
Relativistic corrections are resummed for a class of color-singlet
contributions to all orders in the charm-quark velocity $v_c$ in the
charmonium rest frame. Almost every process that we consider in this
work has negative relativistic corrections ranging from $-20$ to $-35$\,\%.
Among the various processes, the relativistic corrections of
the next-to-leading order in $v_c$ to the decay rate for
$\chi_{b2}\to \eta_c(mS)+\eta_c(nS)$ with $m,$ $n=1$ or 2
are very large. In every case, the resummation of the
relativistic corrections enhances the rate in comparison with the
next-to-leading-order results. We compare our results with
available predictions based on the NRQCD factorization formalism.
The NRQCD predictions are significantly smaller than
those based on the light-cone formalism by 1 or 2 orders
of magnitude.
\end{abstract}
\maketitle


\section{Introduction\label{introduction}}
Among various bottomonium states $H$, the $\eta_b$ and $\chi_{bJ}$ mesons
with $J^{PC}=0^{-+}$ and $1^{++}$, respectively,
have a common feature that the charge conjugation parity $C$ is even.
Because the quantum chromodynamics (QCD) preserves $C$, these mesons
can decay into a pair of charmonia $h_1+h_2$ with the same $C$ parity.
Therefore, possible decay modes of the mesons are
\begin{subequations}
\begin{eqnarray}
\eta_b \textrm{ or } \chi_{bJ}&\to& \psi(mS)+\psi(nS),
\\
\eta_b \textrm{ or } \chi_{bJ}&\to& \eta_c(mS)+\eta_c(nS),
\\
\eta_b \textrm{ or } \chi_{bJ}&\to& \chi_{cJ}+\chi_{cJ'},
\end{eqnarray}
\end{subequations}
where the spin-triplet $S$-wave state $\psi(nS)$ is a $J^{PC}=1^{--}$
eigenstate. However, some of these decay modes like
$\eta_b$ or $\chi_{b1}\to\eta_c(mS)+\eta_c(nS)$  are forbidden
due to the parity conservation in QCD.

Previous theoretical studies include the calculation of the decay
rate for the process $\eta_b\to J/\psi+J/\psi$~\cite{Jia:2006rx}
within the nonrelativistic QCD (NRQCD) factorization
formalism~\cite{Bodwin:1994jh}.
This process was proposed as a candidate for
the discovery mode of the $\eta_b$ meson~\cite{Braaten:2000cm}.
This work was followed by a study of the final-state interaction
in Ref.~\cite{Santorelli:2007xg} and the next-to-leading-order (NLO)
QCD corrections in Refs.~\cite{Gong:2008ue,Sun:2010qx}.
In the case of the spin-triplet $P$-wave decay,
the process $\chi_{bJ}\to J/\psi+J/\psi$ was investigated in
Ref.~\cite{Braguta:2005gw} within the light-cone (LC) formalism.
The authors have extended their LC
predictions to various channels in Ref.~\cite{Braguta:2009df},
where they also provided the NRQCD predictions at leading order (LO)
in the charm-quark velocity $v_c$ in the charmonium rest frame.
According to the results in Ref.~\cite{Braguta:2009df},
the LC predictions are greater than the NRQCD counterparts
although they are in agreement within errors that are significant.
One of the motivations of this work is to investigate if these
discrepancies are reduced under relativistic corrections.\footnote{
Very recently, the authors of Ref.~\cite{Zhang:2011ng} have reported
an NRQCD prediction for the decay $\chi_{bJ}\to J/\psi+J/\psi$
including the relativistic corrections of relative order $v_c^2$, where
they used a velocity-expansion scheme that is
different from the standard NRQCD approach employed in this work.}

In the NRQCD factorization formula, the decay rate is expanded
in powers of the velocity $v_Q$ of the heavy quark $Q$ in
the quarkonium rest frame. In the case of the LC formalism,
the amplitude is expanded in powers of the inverse of the hard scale.
In the limit of $m_b\gg m_c$ as well as $1\gg v_c^2 \gg v_b^2$,
we can guess that the leading-twist LC prediction and the NRQCD
prediction at LO in $v_Q$ are roughly consistent with each other.
However, if there are large relativistic or QCD corrections, such
a naive estimate may fail. For example, the cross section for the
exclusive production process $e^+e^-\to J/\psi+\eta_c$ at the $B$
factories suffers large relativistic~\cite{Bodwin:2007ga} and
QCD~\cite{Zhang:2005cha} corrections. The NRQCD prediction including
relativistic and QCD corrections is shown to be consistent with the
empirical value within errors~\cite{Bodwin:2007ga}. One may guess that
LC predictions as in Ref.~\cite{Bondar:2004sv} provide a reasonable
answer without further corrections. However, as is shown in Ref. [13],
the LC calculation contains
short-distance contributions, which are treated in the context of
model LC distributions. These short-distance contributions appear
in corrections of order $\alpha_s$ (and higher)
in the NRQCD approach and can be computed from first principles in that
approach.
Therefore, it is very interesting to see what happens in the case
of the bottomonium decay into charmonium pairs that have similar
features as the exclusive process $e^+e^-\to J/\psi+\eta_c$.

In this work, we compute the decay rates for the
$C$-even bottomonia  $\eta_b$ and $\chi_{bJ}$ into pairs of
$S$-wave charmonia within the NRQCD factorization formalism.
The computation is carried out within the color-singlet mechanism
of NRQCD because, in these exclusive processes, the color-octet
channel enters only if at least two hadrons involve color-octet
contributions that are suppressed.\footnote{See
Ref.~\cite{Braguta:2009df} for further discussion regarding
the suppression of the color-octet contributions.}
In addition, we also neglect the electromagnetic decay
mode such as $\eta_b\to \gamma^*\gamma^*\to J/\psi+J/\psi$ that
is tiny compared with the QCD mode.\footnote{In the case of
$e^+e^-\to J/\psi+J/\psi$~\cite{Bodwin:2002fk,Bodwin:2002kk},
the LO cross section is comparable to
that~\cite{Braaten:2002fi,Liu:2002wq} of
$e^+e^-\to J/\psi+\eta_c$ because of the enhancements
due to the photon fragmentation and large collinear
emissions from the electron line in the forward region.
In the case of  $\eta_b\to \gamma^*\gamma^*\to J/\psi+J/\psi$
only the photon fragmentation contributes and the collinear
enhancement is missing because of the large bottom-quark mass.
In addition, the fractional electric charge of the bottom quark
makes the rate insignificant. See also Ref~\cite{Jia:2006rx}.}
Relativistic corrections are computed by making use of the
generalized Gremm-Kapustin relation~\cite{Bodwin:2007fz} that has
been employed to resolve the $e^+e^-\to J/\psi+\eta_c$
puzzle~\cite{Bodwin:2007ga}. The method enables us to resum
a class of color-singlet contributions to all orders in $v_Q^2$.
Our calculation reveals that the discrepancies between the NRQCD
prediction and the prediction based on the LC formalism
become more severe, particularly in the $\chi_{b2}$ decay
into an $S$-wave spin-singlet charmonium pair.
This is in contrast to the case of $e^+e^-\to J/\psi+\eta_c$.

This paper is organized as follows.
We first describe the NRQCD factorization formula for
the exclusive $C$-even bottomonium decay into an $S$-wave
charmonium pair in Sec.~\ref{NRQCD-f}.
In Sec.~\ref{Notations} we present the strategy to compute the
short-distance coefficients for the NRQCD factorization formula.
The analytic results for the decay rates at LO and the corrections of
NLO in $v_c^2$ are given in Sec.~\ref{correction-v2}. Our final
numerical results for the decay rates, in which a class of the
color-singlet contributions are resummed to all orders in $v_c^2$,
are listed in Sec.~\ref{correction-vn} and compared with available
predictions. Finally, we summarize the work in Sec.~\ref{summary}.
\section{NRQCD factorization formula\label{NRQCD-f}}
The exclusive decay of a bottomonium into a pair of charmonium states
involves the annihilation of a $b\bar{b}$ pair followed by the creation
of two pairs of $c\bar{c}$. One may guess that the generalization of the NRQCD
factorization \cite{Bodwin:1994jh} for the electromagnetic decay or
light-hadronic decay into this exclusive mode is possible.\footnote{
For example, NRQCD factorization theorems for the exclusive quarkonium
productions in $e^+e^-$ annihilation and $B$ decay have been
proved~\cite{Bodwin:2008nf,Bodwin:2010fi,Bodwin:2009cb}.} If we assume that
the NRQCD factorization is valid for the exclusive decay of a bottomonium
$H$ into a charmonium pair $h_1+h_2$, then the decay rate can be expressed
as a linear combination of the products of nonperturbative NRQCD matrix
elements with numerous spectroscopic states. According to the velocity-scaling
rules of NRQCD~\cite{Bodwin:1994jh}, these matrix elements are classified
in powers of $v_Q$. These exclusive decay modes are dominated by
the color-singlet channels as is stated in the previous section and we
restrict ourselves to the color-singlet contributions. The typical velocity
of the bottom quark $v_b$ in the initial state is significantly smaller than
those of the final-state charmonia, $v_b^2\sim 0.1\ll v_c^2\sim 0.3$
so that we neglect the relativistic effects of the bottom quark while we
include the relativistic corrections of the charm quarks in the
final-state charmonia.

Within the color-singlet mechanism at LO in $v_Q$, there are NRQCD matrix
elements $\langle H|\mathcal{O}_{1}|H\rangle$ for the bottomonium decay
and $\langle 0|\mathcal{O}^{h_i}_{1}|0\rangle$ for the charmonium production
that involve the decay rate $\Gamma[H\to h_1+h_2]$ for $i=1$ and 2.
The spectroscopic states for the four-quark operators $\mathcal{O}_{1}$
and $\mathcal{O}^{h_i}_{1}$ are identical to those of the corresponding
hadrons.\footnote{For the initial-state bottomonia ${}^{2S+1}L_J={}^1S_0$
and ${}^3P_{J}$ with $J=0$, 1, or 2
for $H=\eta_b$ and $\chi_{bJ}$, respectively.
In the charmonium case, ${}^{2S+1}L_J={}^1S_0$ and ${}^3S_1$
for $h_i=\eta_c$ and $J/\psi$, respectively.}
In order to describe the relativistic corrections to the charmonium state,
we denote $\langle 0|\mathcal{O}_{1,(m_i,n_i)}^{h_i}|0\rangle$ as the NRQCD
matrix element for the charmonium production that is of relative order
$v_c^{2(m_i+n_i)}$ in comparison with the LO matrix element
$\langle 0|\mathcal{O}_{1}^{h_i}|0\rangle$. The spectroscopic state
of the four-quark operator $\mathcal{O}_{1,(m_i,n_i)}^{h_i}$ is again
identical to that of $h_i$. As a result, the NRQCD factorization formula
for $\Gamma[H\to h_1+h_2]$ can be expressed as
\begin{equation}
\label{NRQCD-for-1}
\Gamma[H\to h_1+h_2]
=
\langle H|\mathcal{O}_{1}|H\rangle
\sum_{m_i,n_i}
c_{(m_1,n_1),(m_2,n_2)}
\langle 0|\mathcal{O}^{h_1}_{1,(m_1,n_1)}|0\rangle
\langle 0|\mathcal{O}^{h_2}_{1,(m_2,n_2)}|0\rangle,
\end{equation}
where $c_{(m_1,n_1),(m_2,n_2)}$ is the short-distance coefficient
which is insensitive to the long-distance nature of the hadrons
$H$, $h_1$, and $h_2$. These factors can be computed perturbatively
in powers of the strong coupling $\alpha_s$.

The LO color-singlet NRQCD four-quark operator $\mathcal{O}_{1}$
for the annihilation decay of the heavy quarkonium with the
spectroscopic state ${}^{2S+1}L_J$ is of the form
\begin{equation}
\label{b-operator}
\mathcal{O}_{1}=
\psi^\dagger\mathcal{K}({}^{2S+1}L_J)\chi
\chi^\dagger\mathcal{K}({}^{2S+1}L_J)\psi,
\end{equation}
where $\psi$ and $\chi^\dagger$ are the Pauli spinor fields that
annihilate $Q$ and $\bar{Q}$, respectively. For a bottomonium
(charmonium), it is understood to be $Q=b$ ($c$). Here,
the operators $\mathcal{K}({}^{2S+1}L_J)$ are defined by
\begin{subequations}
\label{b-operator-2}
\begin{eqnarray}
\mathcal{K}({}^1S_0)&=&\mathbbm{1},\\
\mathcal{K}^i({}^3S_1)&=&\sigma^i,\\
\mathcal{K}({}^3P_0)&=&\frac{1}{\sqrt{3}}
(-\tfrac{i}{2}\overleftrightarrow{\bm{D}}\cdot\bm{\sigma}),\\
\label{3p1}
\mathcal{K}^i({}^3P_1)&=&\frac{1}{\sqrt{2}}
(-\tfrac{i}{2}\overleftrightarrow{\bm{D}}\times\bm{\sigma})^i,\\
\label{3p2}
\mathcal{K}^{ij}({}^3P_2)&=&
-\tfrac{i}{2}\overleftrightarrow{D}^{(i}\sigma^{j)},
\end{eqnarray}
\end{subequations}
where $\mathbbm{1}$ is the identity matrix for the spin and color space,
$\sigma^i$ is the Pauli matrix and $\bm{D}$ is the gauge-covariant
derivative.
The notation
$A^{(ij)}$ in Eq.~(\ref{3p2}) represents the symmetric traceless component
$\frac{1}{2}(A^{ij}+A^{ji})-\frac{1}{3}A^{kk}\delta^{ij}$
of a Cartesian tensor $A^{ij}$.
The NRQCD matrix element
$\langle H|\mathcal{O}_{1}|H\rangle$
for the decay is averaged over the spin states of $H$.

The NRQCD four-quark operators
$\mathcal{O}_{1,(m,n)}^{h}$ in Eq.~(\ref{NRQCD-for-1})
for the $S$-wave charmonium production can be expressed as
\begin{equation}
\label{ohcab}
\mathcal{O}^{h}_{1,(m,n)}=\frac{1}{2}
\bigg[\chi^\dagger\mathcal{K}_{m}({}^{2S+1}L_J)\psi
\bigg(\sum_{\lambda} a^\dagger_{h_{\lambda}}a_{h_{\lambda}}\bigg)
\psi^\dagger\mathcal{K}_{n}({}^{2S+1}L_J)\chi
+\textrm{H.c.}\bigg],
\end{equation}
where $\mathcal{K}_n$ are defined by
\begin{subequations}
\label{c-operator}
\begin{eqnarray}
\mathcal{K}_n(^1S_0)&=&
(-\tfrac{i}{2}\overleftrightarrow{\bm{\nabla}})^{2n},\\
\mathcal{K}^i_n(^3S_1)&=&
(-\tfrac{i}{2}\overleftrightarrow{\bm{\nabla}})^{2n}\sigma^i.
\end{eqnarray}
\end{subequations}
As is explained earlier, the Pauli spinor fields are for the charm quark.
We have listed only a class of the operators that contain ordinary
derivatives rather than covariant derivatives. The neglect of the
operators with gauge fields contributes first at relative order
$v_c^4$ in the Coulomb gauge in which the matrix elements are
evaluated.\footnote{See Refs.~\cite{Bodwin:2006dn,Bodwin:2008vp}
for further discussion.} Applying the vacuum-saturation approximation,
we can simplify these NRQCD matrix elements as
\begin{equation}
\label{VACSAT}
\langle 0|
\mathcal{O}^{h}_{1,(m,n)}|0\rangle
=(2J+1)
\langle 0|\chi^\dagger\mathcal{K}_{m}\psi|h\rangle
\langle 0|\chi^\dagger\mathcal{K}_{n}\psi|h\rangle^*+{\mathcal O}(v_c^4),
\end{equation}
where $J$ is the total angular momentum of the charmonium
$h$. The spin-multiplicity factor $2J+1$ appears because the spin states
of the produced hadron $h$ are summed over in Eq.~(\ref{ohcab}).

In carrying out the resummation of the relativistic corrections, it is
convenient to define the ratio $\langle\bm{q}^{2n}\rangle_{h}$ of the NRQCD
matrix element of relative order $v_c^{2n}$ to the LO matrix element as
\begin{equation}
\label{ratios}
\langle\bm{q}^{2n}\rangle_{h}
=\frac{\langle 0|\chi^\dagger\mathcal{K}_n\psi|h\rangle}
{\langle 0|\chi^\dagger\mathcal{K}_0\psi|h\rangle}.
\end{equation}
This ratio is independent of the polarization of $h$.
In Ref.~\cite{Bodwin:2006dn} a generalized version of the
Gremm-Kapustin relation \cite{Gremm:1997dq} was derived:
\begin{equation}
\label{relation}
\langle\bm{q}^{2n}\rangle_{h}=\langle\bm{q}^{2}\rangle^n_{h}.
\end{equation}
This relation holds for the matrix elements in spin-independent-potential
models. Thus this relation holds for both spin-singlet and -triplet states
independently of the index $i$ up to corrections of $v_Q^2$ that break
the heavy-quark spin symmetry. This relation has been applied to determine
the NRQCD matrix elements for the $S$-wave quarkonium states
precisely \cite{Bodwin:2007fz,Chung:2010vz} and to resum the relativistic
corrections in various quarkonium
processes \cite{Bodwin:2007ga,Bodwin:2008vp,Lee:2010ts}.
Although there is an intrinsic limitation that the resummation of
relativistic corrections with only the $Q\bar{Q}$ Fock-state contributions
eventually has the predictive power up to corrections of relative order
$v_Q^{4}$~\cite{Bodwin:2007fz}, the method is still useful to improve the
convergence in a process involving significant relativistic corrections.
Such an example is the exclusive $J/\psi+\eta_c$ production in $e^+e^-$
annihilation \cite{Bodwin:2007ga}. We shall find in
Sec.~\ref{correction-vn} that a large correction is indeed observed in
the process $\chi_{b2}\to \eta_c+\eta_c$. Because of the large relativistic
corrections of relative order $v_c^2$, the theoretical prediction for the
cross section can even be negative. This problem will be cured by resumming
the relativistic corrections.

By employing the vacuum-saturation approximation and the generalized
Gremm-Kapustin relation (\ref{relation}), we can express the NRQCD matrix
element for the charmonium production as
\begin{equation}
\langle 0|
\mathcal{O}^{h}_{1,(m,n)}|0\rangle=(2J+1)\langle\bm{q}^2\rangle_{h}^{m+n}
\langle h|
\mathcal{O}_1|h\rangle+O(v_c^4),
\end{equation}
where the errors of order $v_c^4$ are from the vacuum-saturation
approximation. Then the NRQCD factorization formula (\ref{NRQCD-for-1})
for the decay is simplified into the form
\begin{equation}
\label{NRQCD-for-2}
\Gamma[H\to h_1+h_2]
=
\langle H|\mathcal{O}_1|H\rangle
\langle h_1|\mathcal{O}_{1}|h_1\rangle
\langle h_2|\mathcal{O}_{1}|h_2\rangle
\sum_{n_1,n_2}
d_{n_1,n_2}
\langle\bm{q}^{2}\rangle_{h_1}^{n_1}
\langle\bm{q}^{2}\rangle_{h_2}^{n_2},
\end{equation}
where we have redefined the short-distance coefficient as $d_{n_1,n_2}$.
The state $|H\rangle$  in the NRQCD matrix elements has the nonrelativistic
normalization $\langle H(\bm{P})| H(\bm{P}')\rangle=(2\pi)^3$
$\delta^{(3)}(\bm{P}-\bm{P}')$.

The short-distance coefficients $d_{n_1,n_2}$ are determined by the
perturbative matching. In fact, we can construct the NRQCD factorization
formula for the amplitude $\mathcal{A}_{H\to h_1+h_2}$ under the
vacuum-saturation approximation. Once we replace the hadrons $H$, $h_1$,
and $h_2$ with the perturbative heavy-quark-antiquark states $b\bar{b}$,
$c\bar{c}_1$, and $c\bar{c}_2$ with the same spectroscopic states as
those of the corresponding hadrons, respectively, then we find that the
$Q\bar{Q}$ counterpart $\mathcal{A}_{b\bar{b}\to c\bar{c}_1+c\bar{c}_2}$
is calculable perturbatively as
\begin{subequations}
\label{NRQCD-for-3}
\begin{eqnarray}
\label{NRQCD-for-3-H}
\mathcal{A}_{H\to h_1+h_2}
&=&
\sqrt{2m_H}
\sqrt{2m_{h_1}}
\sqrt{2m_{h_2}}
\sqrt{
\langle H|\mathcal{O}_1|H\rangle
\langle h_1|\mathcal{O}_{1}|h_1\rangle
\langle h_2|\mathcal{O}_{1}|h_2\rangle}
\nonumber\\
&&\times
\sum_{n_1,n_2}
a_{n_1,n_2}
\langle\bm{q}^2\rangle^{n_1}_{h_1}
      \langle\bm{q}^2\rangle^{n_2}_{h_2},
\\
\label{NRQCD-for-3-Q}
\mathcal{A}_{b\bar{b}\to c\bar{c}_1+c\bar{c}_2}
&=&
\sqrt{
\langle b\bar{b}|\mathcal{O}_1|b\bar{b}\rangle
\langle c\bar{c}_1|\mathcal{O}_{1}|c\bar{c}_1\rangle
\langle c\bar{c}_2|\mathcal{O}_{1}|c\bar{c}_2\rangle}
\sum_{n_1,n_2}
a_{n_1,n_2}
\bm{q}_1^{2n_1}\bm{q}_2^{2n_2},
\end{eqnarray}
\end{subequations}
where $a_{n_1,n_2}$ is the short-distance coefficient at the amplitude
level and $\bm{q}_i$ is half the relative momentum of the $i$-th
$c\bar{c}$ pair. While the hadron states like $|H\rangle$ and $|h_i\rangle$
are normalized nonrelativistically, we use the relativistic normalization
for the heavy-quark state $|Q\rangle$: $\langle Q(\bm{p})| Q(\bm{p}')\rangle=%
(2\pi)^32(m_Q^2+\bm{p}^2)^{1/2}\delta^{(3)}(\bm{p}-\bm{p}')$.
The normalization factor $\sqrt{8m_H m_{h_1} m_{h_2}}$ was introduced
to make the amplitude $\mathcal{A}_{H\to h_1+h_2}$ have the relativistic
normalization like $\mathcal{A}_{b\bar{b}\to c\bar{c}_1+c\bar{c}_2}$.
Here, $m_H$ and $m_{h_i}$ are the masses of $H$ and $h_i$, respectively.

Because the short-distance coefficients are insensitive to the long-distance
nature of the hadrons, the factor $a_{n_1,n_2}$ in Eq.~(\ref{NRQCD-for-3})
must be common to both $\mathcal{A}_{b\bar{b}\to c\bar{c}_1+c\bar{c}_2}$
and $\mathcal{A}_{H\to h_1+h_2}$. The $Q\bar{Q}$ NRQCD matrix elements
in Eq.~(\ref{NRQCD-for-3-Q}) are calculable perturbatively:
\begin{subequations}
\label{ME-QQbar-norm}
\begin{eqnarray}
\langle b\bar{b}|\mathcal{O}_{1}|b\bar{b}\rangle
&=&2N_c(2E)^2,
\\
\langle c\bar{c}_{i}|\mathcal{O}_{1}|c\bar{c}_{i}\rangle
&=&2N_c(2E_i)^2,
\end{eqnarray}
\end{subequations}
where $E$ and $E_i$ are the energies of the quark in the $Q\bar{Q}$ rest
frame for the $b\bar{b}$ and $c\bar{c}_i$ pairs, respectively.
In such a way, one can determine the short-distance coefficients
$a_{n_1,n_2}$ and $d_{n_1,n_2}$. In the normalization of the
$P$-wave matrix element (\ref{ME-QQbar-norm}) we have suppressed the
factor $|\bm{q}|^2$ that comes from the derivative
operator.\footnote{See, for example, Eq.~(60) of Ref.~\cite{Bodwin:2007zf}.}
\section{Perturbative matching\label{Notations}}
In this section, we present the method to compute the short-distance
coefficients $d_{n_1,n_2}$ in the NRQCD factorization formula for the decay
$H\to h_1+h_2$ by employing the perturbative matching onto the full-QCD
amplitude for $b\bar{b}\to c\bar{c}_1+c\bar{c}_2$.
\subsection{Kinematics}
The momenta for the quarkonia $H$, $h_1$, and $h_2$ are chosen to be
$P$, $P_1$, and $P_2$, respectively. We list the definitions of the variables
for the bottomonium $H$ or the $b\bar{b}$ pair without any index. Variables
with the index $i=1$ or $2$ indicate that they correspond to the charmonium $h_i$
or the $c\bar{c}_i$ pair. For convenience we list the
definitions for various variables for a $Q\bar{Q}$ pair system without any index
unless it is necessary.

We denote $P$ and $q$ by the total and half the relative momentum of a
$Q\bar{Q}$ pair, respectively. Then the momenta for the quark ($p$) and
the antiquark ($\bar{p}$) are expressed as
\begin{subequations}
\label{kinematic-1}
\begin{eqnarray}
p&=&\frac{1}{2}P+q,\\
\bar{p}&=&\frac{1}{2}P-q.
\end{eqnarray}
\end{subequations}
It is obvious, in the rest frame of the $Q\bar{Q}$ pair, that $P$
and $q$ are orthogonal: $P\cdot q=0$. In that frame,
\begin{subequations}
\label{kinematic-2}
\begin{eqnarray}
P&=&(2E, 0),\\
q&=&(0,\bm{q}),\\
p&=&(E, \bm{q}),\\
{\bar p}&=&(E,-\bm{q}),
\end{eqnarray}
\end{subequations}
where $E=\sqrt{m_Q^2+\bm{q}^2}$ is the energy of the
$Q$ or $\bar{Q}$ in the $Q\bar{Q}$ rest frame and $m_Q$
is the mass of the heavy quark $Q$. We assume that the $Q$
and $\bar{Q}$ are on their mass shells so that
\begin{subequations}
\label{kinematic-3}
\begin{eqnarray}
p^2&=&\bar{p}^{\,2}=m_Q^2,\\
P^2&=&4E^2.
\end{eqnarray}
\end{subequations}
In addition, we define the ratio
\begin{equation}
\label{ratio-r}
r\equiv\frac{m_c^2}{m_b^2},
\end{equation}
which is useful to simplify expressions.
\subsection{Spin and color projectors}
\begin{figure}[ht]
\begin{center}
\includegraphics[height=5cm]{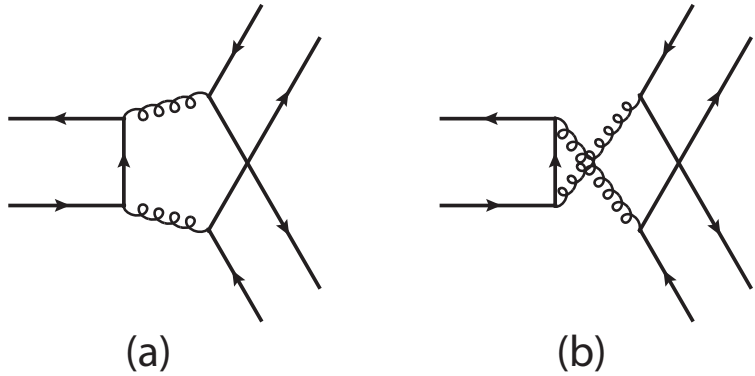}
\caption{The Feynman diagrams for the exclusive decay of
a $C$-even bottomonium decay into a pair of charmonium states
at LO in $\alpha_s$.}
\label{feynman:fig}
\end{center}
\end{figure}

The Feynman diagrams for the process $b\bar{b}\to c\bar{c}_1+c\bar{c}_2$
at LO in $\alpha_s$ are shown in Fig.~\ref{feynman:fig}.
The corresponding perturbative amplitude for the process
$b\bar{b}\to c\bar{c}_1+c\bar{c}_2$ is of the form
\begin{equation}
\mathcal{A}
=\frac{-ig_s^4}{(p_1+\bar{p}_2)^2(p_2+\bar{p}_1)^2}
\bar{v}(\bar{p},\bar{s})\mathcal{B}_{\mu\nu}u(p,s)
\bar{u}(p_1,s_1)\mathcal{C}^\mu v(\bar{p}_2,\bar{s}_2)
\bar{u}(p_2,s_2)\mathcal{C}^\nu v(\bar{p}_1,\bar{s}_1),
\label{amplitude-1}
\end{equation}
where $g_s$ is the strong coupling, the denominator factor is from
the gluon propagators and $\bar{s}$ and ${s}$ ($\bar{s}_i$ and ${s}_i$)
are the spins of $\bar{b}$ and $b$ ($\bar{c}$ and $c$ in the $i$-th
$c\bar{c}$ pair), respectively. $\mathcal{B}^{\mu\nu}$ contains the
bottom-quark propagator and the gluon vertices to the bottom-quark line.
$\mathcal{C}^\mu$ is the gluon vertex to the charm-quark pair $c\bar{c}$
and we have suppressed the color indices. Both $\mathcal{B}^{\mu\nu}$ and
$\mathcal{C}^\mu$ act on spinors with both Dirac and color indices.

In order to compute the amplitude for the perturbative process
$b\bar{b}\to c\bar{c}_1+c\bar{c}_2$ with the appropriate spectroscopic
states, it is convenient to use the projection operators. In this work,
we consider only the color-singlet contributions that can be projected out
by replacing the color component of the outer product of the spinors for
$Q$ and $\bar{Q}$ in each $Q\bar{Q}$ pair with the color-singlet projector
\begin{equation}
\pi_1=\frac{1}{\sqrt{N_c}}\mathbbm{1},
\label{color-pro-1}
\end{equation}
where $\mathbbm{1}$ is the unit color matrix.

The spin-singlet and -triplet components of each $Q\bar{Q}$ state can be
projected out by making use of the spin projectors. After multiplying
corresponding Clebsch-Gordan coefficients to the spin component of the
outer product of the spinors for each $Q\bar{Q}$ pair, one can find the
spin-singlet and -triplet projectors $\Pi_1$ and $\Pi_3$ for the $Q\bar{Q}$
decay and $\bar{\Pi}_1$ and $\bar{\Pi}_3$ for the $Q\bar{Q}$ production,
respectively. The spin projectors that are valid to all orders in the relative
momentum can be found in Refs.~\cite{Bodwin:2002hg,Bodwin:2010fi}.
\begin{subequations}
\label{spin-pro-1}
\begin{eqnarray}
\Pi_1&=&-N
(\,/\!\!\!{p}+m_Q)
(\,/\!\!\!\!P+\!2E)\,\gamma_5
(\,/\!\!\!\bar{p}-m_Q),\\
\Pi_{3\alpha}(\lambda)\epsilon^\alpha(\lambda) &=&\phantom{-}N
(\,/\!\!\!{p}+m_Q)(\,/\!\!\!\!P+\!2E)\,/\!\!\!\epsilon(\lambda)
(\,/\!\!\!\bar{p}-m_Q),\\
\bar{\Pi}_1&=&\phantom{-}N
(\,/\!\!\!\bar{p}-m_Q)\,\gamma_5(\,/\!\!\!\!P+\!2E)\,(\,/\!\!\!{p}+m_Q),\\
\bar{\Pi}_{3\alpha}(\lambda)\epsilon^{*\alpha}(\lambda)&=&\phantom{-}N
(\,/\!\!\!\bar{p}-m_Q)\,/\!\!\!\epsilon^*(\lambda)(\,/\!\!\!\!P+\!2E)\,
(\,/\!\!\!{p}+m_Q),
\end{eqnarray}
\end{subequations}
where the normalization factor $N$ is
\begin{equation}
N=\frac{1}{4\sqrt{2}E(E+m_Q)},
\end{equation}
if we choose the relativistic normalization for the spinors.
$\epsilon(\lambda)$ is the polarization four-vector of the spin-triplet
state with the helicity $\lambda$.
\subsection{Projection of ${S}$- and ${P}$-wave contributions}
We can extract the color-singlet amplitude $\mathcal{A}_1$ with appropriate
spin states of the three quarkonium states from the amplitude $\mathcal{A}$ as
\begin{equation}
\mathcal{A}_1=
\frac{-ig_s^4}{(p_1+\bar{p}_2)^2(p_2+\bar{p}_1)^2}
\textrm{Tr}[\mathcal{B}_{\mu\nu}\Pi|_{b\bar{b}}]\,\,
\textrm{Tr}[\,\mathcal{C}^\mu\,\bar{\Pi}|_{{c\bar{c}}_2}\,
              \mathcal{C}^\nu\,\bar{\Pi}|_{{c\bar{c}}_1}],
\label{amp-spin}
\end{equation}
where $\mathcal{B}_{\mu\nu}$ and $\mathcal{C}^\mu$ are
those in Eq.~(\ref{amplitude-1}) and $\Pi$ ($\bar{\Pi}$) is the direct
product of the color and spin projectors defined in
Eqs.~(\ref{color-pro-1}) and (\ref{spin-pro-1}):
\begin{subequations}
\begin{eqnarray}
\Pi&=&\pi_1\otimes\left(\Pi_{1}\textrm{ or }
\Pi_{3\alpha}\epsilon^\alpha\right),
\\
\bar{\Pi}&=&\pi_1\otimes\left(\Pi_{1}\textrm{ or }
\bar{\Pi}_{3\alpha}\epsilon^{*\alpha}\right).
\end{eqnarray}
\end{subequations}
The trace in Eq.~(\ref{amp-spin}) is over the color and spin indices.

As the next step, we need to pull out the $L$-wave amplitude, where
$L=S$ or $P$, from the color-singlet amplitude $\mathcal{A}_1$ in
Eq.~(\ref{amp-spin}) with correct spin states. In the case of the initial
states $H=\eta_b$ and $\chi_{bJ}$, we need to project out the spin-singlet
$S$-wave and spin-triplet $P$-wave states, respectively. As we have stated
earlier, we consider only the contributions of LO in $v_b$ in the bottomonium
sector. Because the dependence of the momenta $P$ and $q$ for the $b\bar{b}$
pair is isolated in the tensor $B^{\mu\nu}$ in Eq.~(\ref{amp-spin}), the
projection of the $b\bar{b}$ state can be made only with this factor as
\begin{subequations}
\label{amp-b-sector}
\begin{eqnarray}
\label{amp-b-sector-1s0}
\mathcal{B}^{\mu\nu}_{b\bar{b}_1(^{1}S_0)}&=&
\textrm{Tr}\big[\mathcal{B}^{\mu\nu}
\big(\pi_1\otimes \Pi_1|_{b\bar{b}}\big)
\big]\bigg|_{q=0},
\\
\mathcal{B}^{\mu\nu}_{b\bar{b}_1(^{3}P_J)}&=&
\mathcal{P}^{\alpha\beta}_J
\frac{\partial}{\partial q^\alpha}
\textrm{Tr}\big[\mathcal{B}^{\mu\nu}
\big(\pi_1\otimes \Pi_{3\,\beta}|_{b\bar{b}}\big)
\big]\bigg|_{q=0},
\end{eqnarray}
\end{subequations}
where $\mathcal{P}_J^{\mu\nu}$ projects the total angular momentum
state $J$ in the $P$-wave spin-triplet contribution that are defined by
\begin{subequations}
\label{p-wave-pro-1}
\begin{eqnarray}
\mathcal{P}^{\alpha\beta}_0&=&\frac{1}{\sqrt{3}}I^{\alpha\beta},\\
\mathcal{P}^{\alpha\beta}_1&=&\frac{i}{2\sqrt{2}E}
\epsilon^{\alpha\beta\rho\sigma}P_{\rho}\epsilon_\sigma(\lambda),\\
\mathcal{P}^{\alpha\beta}_2&=&\epsilon^{\alpha\beta}(\lambda).
\end{eqnarray}
\end{subequations}
Here, $\epsilon^\sigma(\lambda)$ and $\epsilon^{\alpha\beta}(\lambda)$
are the polarization vector and tensor for the ${}^3P_1$ and ${}^3P_2$
states with the helicity $\lambda$, respectively. In the normalization of
$\mathcal{B}^{\mu\nu}_{b\bar{b}_1(^{3}P_J)}$ in Eq.~(\ref{amp-b-sector}),
we have suppressed the factor $|\bm{q}|$ to make it to be consistent with
the normalization of the matrix element (\ref{ME-QQbar-norm}).
The tensor $I^{\alpha\beta}$ is defined by
\begin{equation}
\label{i0}
I^{\alpha\beta}=
-g^{\alpha\beta}+\frac{P^{\alpha}P^{\beta}}{P^2}.
\end{equation}

Finally, we summarize the method to project out the $S$-wave component of the
amplitude from $\mathcal{A}_1$ in Eq.~(\ref{amp-spin}) for the two-charmonium
final states that are either a spin-singlet or a triplet including relativistic
corrections. Because $\mathcal{A}_1$ in Eq.~(\ref{amp-spin}) is the color-singlet
amplitude of the $c\bar{c}_i$ pairs with appropriate spin states, we only need to
pull out the $S$-wave component that is independent of the direction of
$\bm{q}_i$ but may depend on $\bm{q}_i^2$. We notice that the trace factor that
includes $\mathcal{B}^{\mu\nu}$ depends on both $\bm{q}_1$ and $\bm{q}_2$ through
the bottom-quark propagator. The trace factor for the charm-quark pairs and the
factor of the gluon propagator $1/[(p_1+\bar{p}_2)^2(p_2+\bar{p}_1)^2]$ have the
dependence on both $\bm{q}_1$ and $\bm{q}_2$. After taking the average over the
directions of $\bm{q}_1$ and $\bm{q}_2$, we find our final expression for the
perturbative amplitude $\mathcal{A}_{b\bar{b}\to c\bar{c}_1+c\bar{c}_2}$ as
\begin{equation}
\label{amp-final}
\mathcal{A}_{b\bar{b}\to c\bar{c}_1+c\bar{c}_2}
=-ig_s^4
\int\!\frac{d\Omega_1d\Omega_2}{(4\pi)^2}
\mathcal{B}^{\mu\nu}_{b\bar{b}_1}
\frac{\textrm{Tr}[\,\mathcal{C}_{\mu}\,\bar{\Pi}|_{{c\bar{c}}_2}\,
              \mathcal{C}_{\nu}\,\bar{\Pi}|_{{c\bar{c}}_1}]}
{(p_1+\bar{p}_2)^2(p_2+\bar{p}_1)^2},
\end{equation}
where $\mathcal{B}^{\mu\nu}_{b\bar{b}_1}$ is defined in Eq.~(\ref{amp-b-sector}).
Note the $b\bar{b}$ and $c\bar{c}_i$ pairs in Eq.~(\ref{amp-final})
have definite spectroscopic states that are suppressed. 
Here, $d\Omega_i$ is the solid-angle element that represents the
direction of $\bm{q}_i$.
\subsection{Short-distance coefficients}
The perturbative amplitude $\mathcal{A}_{b\bar{b}\to c\bar{c}_1+c\bar{c}_2}$
depends only on $m_b$, $m_c$, $\bm{q}_i^2$ and polarizations $\lambda$ and
$\lambda_i$ of $b\bar{b}$ and $c\bar{c}_i$ pairs, respectively. By making use
of the normalization of the NRQCD matrix elements for the $Q\bar{Q}$ states
in Eq.~(\ref{ME-QQbar-norm}), we can find the NRQCD factorization formula for the
amplitude of the decay $H(\lambda)\to h_1(\lambda_1)+h_2(\lambda_2)$
\begin{eqnarray}
\label{amp-H-final}
\mathcal{A}_{H\to h_1+h_2}
=
\sqrt{8m_Hm_{h_1}m_{h_2}}
\sqrt{\frac{
\langle H|\mathcal{O}_1|H\rangle
\langle h_1|\mathcal{O}_1|h_1\rangle
\langle h_2|\mathcal{O}_1|h_2\rangle
}{(2E)^2(2E_1)^2(2E_2)^2(2N_c)^3}}
\mathcal{A}_{b\bar{b}\to c\bar{c}_1+c\bar{c}_2}
\Big|_{
\bm{q}_i^2\to \langle \bm{q}^{2}\rangle_{h_i}}\!\!\!,\phantom{xxx}
\end{eqnarray}
where we have suppressed the helicities of the hadrons that are the same
as those for the $Q\bar{Q}$ pairs. Note that $\bm{q}_i^2$ in the perturbative
amplitude $\mathcal{A}_{b\bar{b}\to c\bar{c}_1+c\bar{c}_2}$ is replaced with
the ratio $\langle \bm{q}^{2}\rangle_{h_i}$.

Now we finally find the NRQCD factorization formula for $\Gamma[H\to h_1+h_2]$:
\begin{equation}
\label{NRQCD-for-1-1}
\Gamma[H\to h_1+h_2]
=\int\!\!\frac{d\Phi_2}{(2J+1)2m_H}
\sum_{\lambda,\,\lambda_1,\,\lambda_2}
|\mathcal{A}_{H(\lambda)\to h_1(\lambda_1)+h_2(\lambda_2)}|^2,
\end{equation}
where $J$ is the total angular momentum of $H$, $\int d\Phi_2$ is the
phase space of the final state and the summation is over the polarizations
of the initial and final states. The factor $2m_H$ in the denominator
cancels that of the squared amplitude.\footnote{See Eq.~(\ref{NRQCD-for-3-H}).}
After summing over the polarization states of the particles, the squared
amplitude becomes invariant under rotation and the phase-space integral
becomes trivial
\begin{equation}
\label{phase-space}
\Phi_2=\int\!\!d\Phi_2=\frac{P_{\textrm{CM}}}{4\pi m_H S},
\end{equation}
where  $S$ is the symmetry factor, $S=1$ ($S=2$) for $h_1\neq h_2$
($h_1= h_2)$. $P_{\textrm{CM}}$ is the magnitude of the three-momentum
of $h_i$ in the $H$ rest frame
\begin{equation}
\label{PCM}
P_{\textrm{CM}}=\frac{\lambda^{1/2}(m_H^2,m_{h_1}^2,m_{h_2}^2)}{2m_H},
\end{equation}
where $\lambda(x,y,z)\equiv x^2+y^2+z^2-2(xy+yz+xz)$. Note that we use
the physical masses $m_H$ and $m_{h_i}$ in evaluating the phase space
$\Phi_2$. This prescription respects the physical endpoints of the phase
space without spoiling the gauge invariance.

In evaluating the $S$-wave amplitude (\ref{amp-final}), we follow the
strategy given in Ref.~\cite{Bodwin:2007ga}. In order to investigate the
convergence of the series in $v_{c}^{2n}$, we also provide the fixed-order
prediction at LO and NLO in  $v_c^2$. To compute the fixed-order relativistic
corrections, we first expand the amplitude (\ref{amp-final}) in powers of
$\bm{q}_i$ and then take the angle average of
$\bm{q}_i$-dependent tensors by making use of the following formulas:
\begin{subequations}
\label{amp-spin-swave-v4}
\begin{eqnarray}
\int\!\! \frac{d\Omega_i}{4\pi} q_i^\mu &=&
0,
\\
\int\!\! \frac{d\Omega_i}{4\pi} q_i^\mu q_i^\nu &=&
\frac{\bm{q}_i^2}{3}I_i^{\mu\nu},
\\
\int\!\! \frac{d\Omega_i}{4\pi} q_i^\mu q_i^\nu q_i^\alpha &=&
0,
\\
\int\!\! \frac{d\Omega_i}{4\pi} q_i^\mu q_i^\nu q_i^\alpha q_i^\beta &=&
\frac{\bm{q}_i^4}{15}
\left(
I_i^{\mu\nu}I_i^{\alpha\beta}
+I_i^{\mu\alpha}I_i^{\nu\beta}
+I_i^{\mu\beta}I_i^{\nu\alpha}
\right),
\end{eqnarray}
\end{subequations}
where  $I_i^{\mu\nu}$ is the same as $I^{\mu\nu}$ defined in
Eq.~(\ref{i0}) except that we replace $P$ with $P_i$.

Our final expression for the short-distance coefficient $d_{n_1,n_2}$
in the NRQCD factorization formula (\ref{NRQCD-for-2}) for the decay rate
$\Gamma[H\to h_1+h_2]$ is
\begin{eqnarray}
\label{short-1}
d_{n_1,n_2}=
\frac{4m_{h_1}m_{h_2}\Phi_2}{n_1!n_2!}
\frac{\partial^{n_1}}{\partial\bm{q}_1^{2n_1}}
\frac{\partial^{n_2}}{\partial\bm{q}_2^{2n_2}}
\bigg[\frac{\sum_{\lambda,\lambda_1,\lambda_2}|
\mathcal{A}_{b\bar{b}(\lambda)\to c\bar{c}_1(\lambda_1)+c\bar{c}_2(\lambda_2)}
|^2}
{(2J+1)(2N_c)^3(2m_b)^2(2E_1)^2(2E_2)^2}
\bigg]\bigg|_{\bm{q}_1^2=\bm{q}_2^2=0}.
\phantom{xxx}
\end{eqnarray}
In the summation of the polarization states for the spin-1 and -2 quarkonia,
we use the following formulas:
\begin{subequations}
\begin{eqnarray}
\sum_{\lambda}
\epsilon^\alpha(\lambda)
\epsilon^{*\mu}(\lambda)&=&I^{\alpha\mu},
\\
\sum_{\lambda}
\epsilon^{\alpha\beta}(\lambda)
\epsilon^{*\mu\nu}(\lambda)&=&
\frac{1}{2}(
I^{\alpha\mu}
I^{\beta\nu}
+
I^{\alpha\nu}
I^{\beta\mu}
)-\frac{1}{3}
I^{\alpha\beta}
I^{\mu\nu},
\end{eqnarray}
\end{subequations}
where $I^{\mu\nu}$ is defined in Eq.~(\ref{i0}). For the spin-1 and -2 charmonium
$h_i$, $I^{\mu\nu}$ must be replaced with $I_i^{\mu\nu}$.
\section{The decay rate up to NLO in $v_c^2$}\label{correction-v2}
Now it is straightforward to compute the decay rate based on the strategy and
techniques described in Sec.~\ref{Notations}. In this section, we present
the analytic expressions for the decay rates $\Gamma[H\to h_1+h_2]$ for
$H=\eta_b$, $\chi_{b0}$, $\chi_{b2}$ and for $h_i=\psi(mS)$, $\eta_c(mS)$ up
to NLO in $v_c^2$. In the case of $\chi_{b1}\to J/\psi+J/\psi$, the LO
contribution first appears at order $v_c^4$ and the decay rate is of order
$v_c^8$. This process is so strongly suppressed that we present only the LO
contribution.

We write the decay rate of the form
\begin{equation}
\label{gammaR}
\Gamma[H\to h_1+h_2]=
\Gamma^{\textrm{LO}}[H\to h_1+h_2](1+R),
\end{equation}
where $\Gamma^{\textrm{LO}}$ is the LO contribution and $R$ is the ratio of
the relativistic corrections to the LO contribution. We also define $R_2$
as the corresponding ratio at NLO in $v_c^2$. In the following, we use the
symbols $\psi_i$ and $\eta_i$ for $h_i=\psi(mS)$ and $\eta_c(mS)$, respectively.
\subsection{$\eta_b\to \psi_1+\psi_2$}
Our analytic results for
$\Gamma^{\textrm{LO}}$ and $R_2$ for $\eta_b\to\psi_1+\psi_2$ are given by
\begin{subequations}
\label{rate-NLO-e-p}
\begin{eqnarray}
\Gamma^{\rm LO}[\eta_b\to\psi_1+\psi_2]&=&\frac{4096\pi^3
\alpha_s^4(1-4r)^{3/2}
  }{6561 m_b^{12}r S}
\langle \eta_b |\mathcal{O}_1|\eta_b\rangle
\langle \psi_1 |\mathcal{O}_1|\psi_1\rangle
\langle \psi_2 |\mathcal{O}_1|\psi_2\rangle
\nonumber\\
&&\times
\big(\langle \bm{q}^2\rangle_{\psi_1}+\langle
   \bm{q}^2\rangle_{\psi_2}\big)^2,
\\
R_2[\eta_b\to\psi_1+\psi_2]&=&
\frac{-1}{15 m_b^2r(1-4 r)(\langle \bm{q}^2\rangle_{\psi_1}+\langle
   \bm{q}^2\rangle_{\psi_2})}
   \bigg[3(\langle \bm{q}^2\rangle_{\psi_1}^2+
   \langle \bm{q}^2\rangle_{\psi_2}^2)
   (3+r+8r^2)
\nonumber\\
&&\hbox{}-5\langle \bm{q}^2\rangle_{\psi_1}
\langle \bm{q}^2\rangle_{\psi_2}
(1-18r-16r^2)\bigg],
\end{eqnarray}
\end{subequations}
where the symmetry factor $S=2$ for $\psi_1=\psi_2$ and, otherwise, $S=1$.
The variable $r$ is defined in Eq.~(\ref{ratio-r}). For simplicity,
we have put $m_{h_i}=2E_i$ in Eq.~(\ref{rate-NLO-e-p}) including the
phase-space factor $\Phi_2$. However, when we present the numerical
results, we will use the expression (\ref{PCM}) to evaluate $\Phi_2$.
As shown in Eq.~(\ref{rate-NLO-e-p}),
$\Gamma^{\rm LO}[\eta_b\to\psi_1+\psi_2]$ is of order $v_c^4$.
In the color-singlet spin-singlet $b\bar{b}$ contribution
in Eq.~(\ref{amp-b-sector-1s0}), the Dirac trace
of $\textrm{Tr}[\mathcal{B}_{\mu\nu}\Pi|_{b\bar{b}}]$ in Eq.~(\ref{amp-spin})
must be proportional to $\epsilon^{\mu\nu\alpha\beta}k_{1\alpha}k_{2\beta}$,
where $k_1=p_1+\bar{p}_2$ and $k_2=p_2+\bar{p}_1$ are the momenta for the
virtual gluons. In the limit $v_c=0$, however, the two momenta become
identical to each other $k_1=k_2=(P_1+P_2)/2=(m_b,0,0,0)$, and therefore
the amplitude vanishes before the vector indices are contracted to
the polarization four-vectors of $\psi_1$ and $\psi_2$. As a result, the
leading contribution to the amplitude begins at order $v_c^2$ and
$\Gamma^{\rm LO}[\eta_b\to\psi_1+\psi_2]$ is of order $v_c^4$.
The expression for $\Gamma^{\rm LO}[\eta_b\to\psi_1+\psi_2]$ in
Eq.~(\ref{rate-NLO-e-p}) agrees with Eq.~(22) of Ref.~\cite{Jia:2006rx}.
The result for $R_2$ is new.
\subsection{$\chi_{b0}\to \eta_1+\eta_2$}
The results for $\chi_{b0}\to \eta_1+\eta_2$ are given by
\begin{subequations}
\label{rate-NLO-c-e}
\begin{eqnarray}
\Gamma^{\rm LO}[\chi_{b0}\to\eta_1+\eta_2]&=&
\frac{2048\pi^3 \alpha_s^4(1+2r)^{2}\sqrt{1-4r}}
     {2187 m_b^{10}r S}
\langle \chi_{b0} |\mathcal{O}_1|\chi_{b0} \rangle
\langle \eta_1 |\mathcal{O}_1|\eta_1\rangle
\langle \eta_2 |\mathcal{O}_1|\eta_2\rangle, \phantom{xxxxx}
\\
R_2[\chi_{b0}\to\eta_1+\eta_2]&=&
-\frac{2(1-3r+3 r^2+8r^3)
       \big( \langle \bm{q}^2\rangle_{\eta_1}
            +\langle \bm{q}^2\rangle_{\eta_2}\big)
      }{3m_b^2r \left(1-2r-8 r^2\right)}.
\end{eqnarray}
\end{subequations}
While the amplitude for $\eta_{b}\to\psi_1+\psi_2$ begins at order $v_c^2$,
the LO contribution to the amplitude
$\mathcal{A}_{\chi_{b0}\to \eta_1+\eta_2}$ begins at relative order $v_c^0$.
The analytic expression for $\Gamma^{\rm LO}$ was also given
in Ref.~\cite{Braguta:2009df}, where the authors evaluated the phase-space
factor $\Phi_2$ in the limit of $m_{h_i}\to 0$. Aside from this difference
in the phase-space factor, the analytic expression for
$\Gamma^{\textrm{LO}}[\eta_{b}\to\eta_1+\eta_2]$ in Eq.~(\ref{rate-NLO-c-e})
is smaller than that in Ref.~\cite{Braguta:2009df} by a factor of 2.
\subsection{$\chi_{b2}\to \eta_1+\eta_2$}
The results for $\chi_{b2}\to \eta_1+\eta_2$ are given by
\begin{subequations}
\label{rate-NLO-c2-e}
\begin{eqnarray}
\Gamma^{\rm LO}[\chi_{b2}\to\eta_1+\eta_2]&=&
\frac{1024\pi^3 \alpha_s^4(1-4r)^{5/2}}
     {10935 m_b^{10}r S}
\langle \chi_{b2} |\mathcal{O}_1|\chi_{b2} \rangle
\langle \eta_1 |\mathcal{O}_1|\eta_1\rangle
\langle \eta_2 |\mathcal{O}_1|\eta_2\rangle, \phantom{xxx}
\\
R_2[\chi_{b2}\to\eta_1+\eta_2]
&=&-\frac{(7+10 r-32 r^2)
   \big(\langle \bm{q}^2\rangle_{\eta_1}+
   \langle \bm{q}^2\rangle_{\eta_2}\big)}
   {6m_b^2 r \left(1-4 r\right)}.
\end{eqnarray}
\end{subequations}
As in the case of $\Gamma^{\textrm{LO}}[\eta_{b}\to\eta_1+\eta_2]$,
$\Gamma^{\rm LO}[\chi_{b2}\to\eta_1+\eta_2]$ in Eq.~(\ref{rate-NLO-c2-e})
is smaller than the corresponding result in Ref.~\cite{Braguta:2009df}
by a factor of 2. The result for $R_2$ is new.
\subsection{$\chi_{b0}\to \psi_1+\psi_2$}
The LO decay rate and relativistic correction factor $R_2$
for $\chi_{b0}\to \psi_1+\psi_2$ are
\begin{subequations}
\label{rate-NLO-c0-p}
\begin{eqnarray}
\Gamma^{\rm LO}[\chi_{b0}\to\psi_1+\psi_2]&=&
\frac{2048\pi^3 \alpha_s^4(1-4r+12r^2)\sqrt{1-4r}
  }{2187m_b^{10} r S}
\langle \chi_{b0} |\mathcal{O}_1|\chi_{b0} \rangle
\nonumber\\
&&\times
\langle \psi_1 |\mathcal{O}_1|\psi_1\rangle
\langle \psi_2 |\mathcal{O}_1|\psi_2\rangle,
\\
R_2[\chi_{b0}\to\psi_1+\psi_2]
&=&-\frac{2(2-11r+14r^2+16r^3)
   \big(\langle \bm{q}^2\rangle_{\psi_1}+
   \langle \bm{q}^2\rangle_{\psi_2}\big)}
   {m_b^2(1-8r+28r^2-48r^3)}.
\end{eqnarray}
\end{subequations}
The results in Eq.~(\ref{rate-NLO-c0-p}) agree with those
Ref.~\cite{Zhang:2011ng}.\footnote{Since the authors in Ref.~\cite{Zhang:2011ng}
adopted a different relativistic expansion, the comparison has been made
after expanding $m_{J/\psi}=2\sqrt{m_c^2+{\bm q}^2}$ in their expressions
in powers of $\bm{q}^2$.}
Like $\Gamma^{\textrm{LO}}[\eta_{b}\to\eta_1+\eta_2]$ and
$\Gamma^{\rm LO}[\chi_{b2}\to\eta_1+\eta_2]$,
$\Gamma^{\rm LO}[\chi_{b0}\to\psi_1+\psi_2]$ in Eq.~(\ref{rate-NLO-c0-p})
is smaller than the corresponding result in Ref.~\cite{Braguta:2009df}
by a factor of 2. The result for $R_2$ is new.
\subsection{$\chi_{b1}\to \psi_1+\psi_2$}
The LO amplitude for $\chi_{b1}\to \psi_1+\psi_2$ is of order $v_c^4$ to make
the decay rate of order $v_c^8$. Because the process is highly suppressed, we only
present the decay rate at the LO in $v_c$:
\begin{eqnarray}
\label{rate-NLO-c1-p}
\Gamma^{\rm LO}[\chi_{b1}\to\psi_1+\psi_2]&=&
\frac{512 \pi ^3 (1-4 r)^{5/2}\alpha_s^4}{4428675 m_b^{18} r^4 S}
\langle \chi_{b1} |\mathcal{O}_1|\chi_{b1} \rangle
\langle \psi_1 |\mathcal{O}_1|\psi_1\rangle
\langle \psi_2 |\mathcal{O}_1|\psi_2\rangle \phantom{xxx}\nonumber\\
&\times&\big[
9r(8 r+25)(\langle \bm{q}^2\rangle_{\psi_1}^4
          +\langle \bm{q}^2\rangle_{\psi_2}^4)
+480r (r+2)\langle \bm{q}^2\rangle_{\psi_1}
           \langle \bm{q}^2\rangle_{\psi_2}
\nonumber\\
&\times&
(\langle \bm{q}^2\rangle_{\psi_1}^2
+\langle \bm{q}^2\rangle_{\psi_2}^2)
+2 (472 r^2+1375 r+1600)\langle \bm{q}^2\rangle_{\psi_1}^2
                        \langle \bm{q}^2\rangle_{\psi_2}^2
\big].\phantom{xxx}
\end{eqnarray}
The expression $\Gamma^{\rm LO}[\chi_{b1}\to\psi_1+\psi_2]$
in Eq.~(\ref{rate-NLO-c1-p}) is new.
\subsection{$\chi_{b2}\to \psi_1+\psi_2$}
The LO decay rate and relativistic correction factor $R_2$
for $\chi_{b2}\to \psi_1+\psi_2$ are
\begin{subequations}
\label{rate-NLO-c2-p}
\begin{eqnarray}
\Gamma^{\rm LO}[\chi_{b2}\to\psi_1+\psi_2]&=&
\frac{1024\pi^3 \alpha_s^4(13+56r+48r^2)\sqrt{1-4r}
  }{10935
   m_b^{10}r S}\langle \chi_{b2} |\mathcal{O}_1|\chi_{b2} \rangle
\nonumber\\
&&\times
\langle \psi_1 |\mathcal{O}_1|\psi_1\rangle
\langle \psi_2 |\mathcal{O}_1|\psi_2\rangle, \phantom{xxx}
\\
R_2[\chi_{b2}\to\psi_1+\psi_2]
&=&-\frac{(13+62r-32r^2-608r^3-512r^4)
   (\langle \bm{q}^2\rangle_{\psi_1}+
   \langle \bm{q}^2\rangle_{\psi_2})}
   {m_b^2r(26+8r-352r^2-384r^3)}.\phantom{xxx}
\end{eqnarray}
\end{subequations}
While the results in Eq.~(\ref{rate-NLO-c2-p}) are in
agreement with those in Ref.~\cite{Zhang:2011ng},
$\Gamma^{\rm LO}[\chi_{b2}\to\psi_1+\psi_2]$ is smaller than the
corresponding result in Ref.~\cite{Braguta:2009df} by a
factor of 2.
\section{Resummation of relativistic corrections and numerical results}
\label{correction-vn}
In Sec.~\ref{correction-v2}, we have listed the NRQCD factorization formulas
for the decay rates $\Gamma[H\to h_1+h_2]$ for various processes, in which
the LO predictions and the NLO corrections with respect to $v_c^{2}$ are
included. In this section, we provide our predictions for these decay rates
including relativistic corrections to all orders in $v_c^{2}$ in which
a class of color-singlet contributions are resummed. Still we neglect $v_b$.
The approach that was employed in Sec.~\ref{correction-v2} at fixed orders
in $v_c^{2}$ cannot be used to resum the relativistic corrections to all
orders in $v_c^{2n}$. Instead of carrying out the $v_c^{2}$ expansion
order by order, we evaluate the average of the amplitude (\ref{amp-final})
over the direction of $\bm{q}_i$ numerically following a previous analysis
in Sec. IV of Ref.~\cite{Bodwin:2007ga}.\footnote{In some simple cases,
complete analytic expressions that contain the relativistic corrections
resummed to all orders in $v_c^{2}$ are known. See, for example,
Refs.~\cite{Bodwin:2008vp,Lee:2010ts,Chung:2010vz}.} Because the
calculations are carried out for the perturbative amplitude, expressions
are for the parton process $b\bar{b}\to c\bar{c}_1+c\bar{c}_2$ rather
than the process $H\to h_1+h_2$.

For simplicity, we carry out the calculation in the rest frame of the
initial $b\bar{b}$ pair. In this frame the explicit components of the
four-momenta for the $b\bar{b}$, $c\bar{c}_1$, and $c\bar{c}_2$ are
given by
\begin{subequations}
\label{momenta-choice}
\begin{eqnarray}
P^*&=&(2m_b,0,0,0),\\
P_1^*&=&(\tilde{E}_{1},0,0,\phantom{-}P_{\textrm{CM}}),\\
P_2^*&=&(\tilde{E}_{2},0,0,-P_{\textrm{CM}}),
\end{eqnarray}
\end{subequations}
where the superscript in a four-vector $V^*$ indicates that
the four-vector is defined in the $b\bar{b}$ rest frame,
$P_{\textrm{CM}}$ and $\tilde{E}_{i}$ are the momentum
and the energy of the $c\bar{c}_i$ pair:
\begin{subequations}
\begin{eqnarray}
\label{kine-1}
P_{\textrm{CM}}&=&\frac{\lambda^{1/2}[(2m_b)^2,(2E_1)^2,(2E_2)^2]}{4m_b},\\
{}\tilde{E}_{i}&=&\sqrt{(2E_i)^2+P_{\textrm{CM}}^2}.
\end{eqnarray}
\end{subequations}

We can parametrize the momentum $q_i$ in the rest frame of
the $c\bar{c}_i$ rest frame as
\begin{equation}
q_i=|\bm{q}_i|(0,\sin\theta_i\cos\phi_i,\sin\theta_i\sin\phi_i,\cos\theta_i),
\end{equation}
where $\theta_i$ and $\phi_i$ are the polar and azimuthal angles of
$\bm{q}_i$ in the $c\bar{c}_i$ rest frame. After
boosting $q_1$ and $q_2$ to the $b\bar{b}$ rest frame, we find that
\begin{subequations}
\begin{eqnarray}
\label{momenta-qi}
q_1^*&=&|\bm{q}_1|(+\gamma_1\beta_1\cos\theta_1,
\sin\theta_1\cos\phi_1,\sin\theta_1\sin\phi_1,\gamma_1\cos\theta_1),\\
q_2^*&=&|\bm{q}_2|(-\gamma_2\beta_2\cos\theta_2,
\sin\theta_2\cos\phi_2,\sin\theta_2\sin\phi_2,\gamma_2\cos\theta_2),
\end{eqnarray}
\end{subequations}
where
\begin{subequations}
\begin{eqnarray}
\gamma_i&=&\frac{1}{\sqrt{1-\beta_i^2}},\\
\beta_i&=&\frac{P_{\textrm{CM}}}{\tilde{E}_i}.
\end{eqnarray}
\end{subequations}
In evaluating the angle average of the amplitude, we have to choose the
numerical values for the input parameters such as the strong coupling
$\alpha_s$, heavy-quark mass $m_Q$, NRQCD matrix elements and physical
masses $m_H$ and $m_{h_i}$ of the hadrons. In this work, we take
$\alpha_s(m_b)=0.215$ that is, within uncertainties, consistent with
previous analyses on
$\chi_{bJ}\to c\bar{c}+X$~\cite{Bodwin:2007zf},
$\Upsilon(1S)\to c\bar{c}+X$~\cite{Kang:2007uv}, and
$\eta_b\to\gamma\gamma$~\cite{Chung:2010vz}. The masses of the heavy quarks
$c$ and $b$ are chosen to be the one-loop pole masses $m_c=1.4$\,GeV and
$m_b=4.6$\,GeV, respectively, following previous NRQCD
analyses~\cite{Braaten:2002fi,Bodwin:2002fk,Bodwin:2007zf,Kang:2007uv,%
Chung:2010vz}.
The most recent fit for the matrix element for $\eta_b(1S)$
can be found in Ref.~\cite{Chung:2010vz}, which is consistent, within
uncertainties, with the values used in Ref.~\cite{Braguta:2009df}.
That of the $P$-wave states $\chi_{bJ}$ can be found in
Ref.~\cite{Bodwin:2007zf}. In the case of $J/\psi$, the color-singlet
NRQCD matrix elements were fit to the electromagnetic decay rate of the state
in which the relativistic corrections are resummed to all orders in $v_c^{2n}$
including the order-$\alpha_s$ corrections. These values can be found in
Ref.~\cite{Bodwin:2007fz}. For the spin-singlet states $\eta_c(1S)$ and
$\eta_c(2S)$, we quote the recent values in Ref.~\cite{Guo:2011tz}
that were fitted to the light-hadronic and electromagnetic decay rates
with the NRQCD factorization formula with the accuracies of order
$\alpha_sv_c^2$.
All of these NRQCD matrix elements with corresponding references
are listed in Table.~\ref{table1}. The values for the physical masses
for the involving hadrons are taken from Ref.~\cite{Nakamura:2010zzi}
as $m_{\eta_{b}}=9.3909\,\textrm{GeV}$,
$m_{\chi_{b0}}=9.85944\,\textrm{GeV}$,
$m_{\chi_{b1}}=9.89278\,\textrm{GeV}$,
$m_{\chi_{b2}}=9.91221\,\textrm{GeV}$,
$m_{\eta_c}=2.9803\,\textrm{GeV}$,
$m_{J/\psi}=3.096916\,\textrm{GeV}$, and
$m_{\eta_c(2S)}=3.637\,\textrm{GeV}$.

\begin{table}
\centering
\caption{
The color-singlet NRQCD matrix elements (ME)
$\langle h| \mathcal{O}_1|h\rangle$ that is LO in $v_Q$
and the ratio $\langle \bm{q}^2\rangle_h$
for various heavy quarkonia $h$. We keep only two digits of significant
figures.
\label{table1}}
\begin{tabular}{ccccccc}
\hline
ME $\backslash$ $h$ & $\eta_b$~\cite{Braguta:2009df}
&$\chi_{bJ}$~\cite{Bodwin:2007zf}&
$\eta_c$~\cite{Guo:2011tz}&$\eta_c(2S)$~\cite{Guo:2011tz}
&$J/\psi$~\cite{Bodwin:2007fz}\\
\hline
$\langle\mathcal{O}^{0} \rangle_h$&$3.1\,{\rm GeV}^3$&$2.0\,{\rm GeV}^5$
&$0.40\,{\rm GeV}^3$&$0.20\,{\rm GeV}^3$&$0.44\,{\rm GeV}^3$\\
\hline
$\langle\bm{q}^2 \rangle_h$&--&--&$0.45\,{\rm GeV}^{2}$&$0.50\,{\rm GeV}^{2}$&
$0.44\,{\rm GeV}^{2}$\\
\hline
\end{tabular}
\end{table}

\begin{table}
\centering
\caption{
The decay rate for the process $H\to h_1+h_2$. $\Gamma^{\textrm{LO}}$ is
the NRQCD prediction at LO in $v_b$ and $v_c$. $\Gamma$ is that in which
the relativistic corrections of all orders in $v_c^{2n}$ are resummed.
The relativistic correction factor $R$ defined in Eq.~(\ref{gammaR}) resummed
to all orders in $v_c^{2n}$ and $R_2$ at NLO order in $v_c^{2}$.
$\Gamma^{\rm LC}$ represents the LC prediction based 
on the formula given in Ref.[8]. $\Gamma^{\textrm{LO}}$, $\Gamma$ and
$\Gamma^{\textrm{LC}}$ are in units of eV.
\label{table2}}
\begin{tabular}{lcccccc}
\hline
${\rm channel}$&$\Gamma^{\rm LO}$& $\Gamma$ & $R_2$
& $R$ & $\Gamma^{\rm LC}$&\\
\hline
$\eta_b\to J/\psi+J/\psi$&
0.58&0.45&
$-27$\%&
$-22$\%&
--&\\
\hline
$\chi_{b0}\to \eta_c+\eta_c$&28&21&$-31$\%&$-25$\%&51&\\
$\chi_{b0}\to \eta_c+\eta_c(2S)$&28&23&$-33$\%&$-17$\%&76&\\
$\chi_{b0}\to \eta_c(2S)+\eta_c(2S)$&7.0&6.4&$-34$\%&$-10$\%&30&\\
\hline
$\chi_{b2}\to \eta_c+\eta_c$&0.79&0.32&$-93$\%&$-59$\%&16&\\
$\chi_{b2}\to \eta_c+\eta_c(2S)$&0.79&0.35&$-98$\%&$-56$\%&21&\\
$\chi_{b2}\to \eta_c(2S)+\eta_c(2S)$&0.20&0.09&$-103$\%&$-54$\%&8&\\
\hline
$\chi_{b0}\to J/\psi+J/\psi$&18&15&$-20$\%&$-16$\%&79&\\
\hline
$\chi_{b1}\to J/\psi+J/\psi$&$1.0\times 10^{-3}$&$3.1\times10^{-4}$&--&$-70$\%&--&\\
\hline
$\chi_{b2}\to J/\psi+J/\psi$&45&35&$-34$\%&$-23$\%&270&\\
\hline
\end{tabular}
\end{table}

Our NRQCD predictions for the decay rate $\Gamma[H\to h_1+h_2]$, in which
the relativistic corrections are resummed to all orders in $v_c^{2n}$
keeping $v_b=0$, are listed in Table.~\ref{table2} in units of eV.
Here, $\Gamma^{\textrm{LO}}$ is the prediction at LO in $v_b$ and $v_c$.
In order to demonstrate the significance of the resummation, we list the
values for $R$ and $R_2$ that are defined in Eq.~(\ref{gammaR}).
$\Gamma^{\textrm{LC}}$ is the prediction based on the LC formula
quoted from Ref.~\cite{Braguta:2009df}. The relativistic corrections to
all of the processes listed in Table.~\ref{table2} are negative. According
to $R_2$ in Table.~\ref{table2}, the NLO relativistic corrections are
between $-20$ and $-35$\,\% of the LO prediction except for
$\chi_{b2}$ decays into spin-singlet $S$-wave charmonium pairs
$\eta_1+\eta_2$ that are more than $-90$\,\%. Especially, the decay
rate for $\chi_{b2}\to \eta_c(2S)+\eta_c(2S)$ becomes negative at NLO
in $v_c^2$. The resummed results for the decay rates are greater than the
NLO-corrected values. The resummation of the relativistic
corrections to all orders in $v_c^{2n}$ makes all of the rates
$\Gamma[\chi_{b2}\to \eta_1+\eta_2]$ positive. The previous predictions
of the decay rates $\Gamma^{\textrm{LC}}$ in Ref.~\cite{Braguta:2009df}
based on the LC formula are greater than our final results $\Gamma$
that include resummation of relativistic corrections to all orders in
$v_c^{2n}$.\footnote{While we list only the central values of our
predictions in Table.~\ref{table2}, the authors of Ref.~\cite{Braguta:2009df}
have provided the results with uncertainties that are huge. The
discrepancies between NRQCD and LC predictions may be partially relieved
under those uncertainties.} For $\chi_{b0}\to \eta_1+\eta_2$, the LC
results are greater than ours by factors ranging from 2 to 5. Especially,
in the case of $\chi_{b2}\to \eta_1+\eta_2$, the LC results are greater
than ours by factors ranging from 45 to 90. In the case of
$\chi_{bJ}\to J/\psi+J/\psi$ the factors are 5 and 8 for $J=0$ and 2.
\section{Summary}
\label{summary}
We have presented the NRQCD predictions for the decay rates of
the bottomonia $\eta_b$ and $\chi_{bJ}$, that are even
eigenstates of the charge conjugation parity $C$, into $S$-wave charmonium pairs.
The short-distance coefficients for the NRQCD factorization formula are
obtained at LO in $\alpha_s$. A class of relativistic corrections of the charm
quark in the final-state charmonia is resummed to all orders in $v_c^{2n}$ by
making use of the generalized Gremm-Kapustin relation in
Refs.~\cite{Bodwin:2006dn,Bodwin:2007fz} that is valid in spin-independent
potential models and we have neglected the motion of the $b$ quark in the
initial bottomonium.

The results show that the relativistic corrections to the decay rates at
NLO in $v_c^2$ are all negative. Severe relativistic corrections
are observed especially in $\chi_{b2}\to\eta_1+\eta_2$ for $\eta_i=\eta_c$ or
$\eta_c(2S)$. The decay rate for $\chi_{b2}\to \eta_c(2S)+\eta_c(2S)$
at NLO in $v_c^2$ is even negative. In almost every case, the
resummation of relativistic corrections of a class of color-singlet
contributions eventually gives sizable growth of the decay rate in
comparison with the NLO predictions. Parts of our results are in
agreement with previous NRQCD predictions in Refs.~\cite{Jia:2006rx,%
Zhang:2011ng}.

In comparison with a previous analysis based on the LC formalism
in Ref.~\cite{Braguta:2009df}, our NRQCD results severely underestimate
the decay rates by 1 or 2 orders of magnitude. It is, therefore,
very important to pin down the source of such significant discrepancies
between the NRQCD factorization and LC formalisms. We recall previous
theoretical studies on the exclusive process $e^+e^-\to J/\psi+\eta_c$
at $B$ factories. The NRQCD predictions of the cross section at LO in
$\alpha_s$ and $v_c$~\cite{Braaten:2002fi,Liu:2002wq} severely
underestimated the empirical values~\cite{Abe:2002rb,Abe:2004ww,%
Aubert:2005tj}. According to Ref.~\cite{Bondar:2004sv},
the LC prediction, which is much larger than that of NRQCD at
LO in $\alpha_s$ and $v_c$, can explain the measured cross section.
However, as is shown in Ref. [13], the LC calculation contains
short-distance contributions, which are treated in the context of
model LC distributions. These short-distance contributions appear
in corrections of order $\alpha_s$ (and higher)
in the NRQCD approach and can be computed from first principles in that
approach. We know that the
problem has been resolved within the NRQCD factorization formula in
combination with the NLO corrections in $\alpha_s$~\cite{Zhang:2005cha}
and the resummation of relativistic corrections~\cite{Bodwin:2007ga}
within errors.
\begin{acknowledgments}
We thank Geoff Bodwin for his critical reading the manuscript.
U.K. and J.L. and W.S.  were supported by Basic Science Research
Program through the NRF of Korea funded by the MEST under Contract
No. 2010-0028228. J.L. and W.S. were also supported in part by a Korea
University fund. The research of R.R. was supported by the National
Natural Science Foundation of China under Grants No.~10875130, 10935012.
\end{acknowledgments}

\end{document}